\documentclass[preprint,prb,aps,eqsecnum,graphicx]{revtex4}
\usepackage{graphicx}
\begin{document}

\title{Effects of impurity scattering on
electron-phonon resonances in semiconductor superlattice
high-field transport }

\author{Shaoxin Feng$^{ab}$, Christoph H. Grein$^a$, and Michael
E. Flatt\'{e}$^b$}

\affiliation{ $^a$Department of Physics, University of Illinois at
Chicago, Chicago,IL 60607\\ $^b$Department of Physics and
Astronomy, University of Iowa, Iowa City, IA, 52242}

\begin{abstract}
A non-equilibrium Green's function method is applied to model
high-field quantum transport and electron-phonon resonances in
semiconductor superlattices. The field-dependent density of states
for elastic (impurity) scattering is found non-perturbatively in
an approach which can be applied to both high and low electric
fields. $I-V$ curves, and specifically electron-phonon resonances,
are calculated by treating the inelastic (LO phonon) scattering
perturbatively.  Calculations show how strong impurity scattering
suppresses the electron-phonon resonance peaks in $I-V$ curves,
and their detailed sensitivity to the size, strength and
concentration of impurities.
\end{abstract}
\maketitle

\section{Introduction}
\label{intro}

There has been tremendous interest in the electron transport
properties of semiconductor superlattices (SLs) since the
pioneering work of Esaki and Tsu\cite{Esaki,Esaki2}. Bragg
scattering of carriers from the Brillouin zone boundaries of a
crystal produces Bloch oscillations, or, equivalently, localized
(Wannier-Stark) states within each unit cell period if the system
is sufficiently clean. This regime is characterized by negative
differential conductance (NDC) at sufficiently high fields, and
occurs if the Bloch frequency $\Omega=eFL/\hbar$ ($F$ is the
electric field and $L$ the unit cell period) is larger than the
effective scattering rate $1/\tau_{eff}$.  The large unit cell of
the superlattice can allow electrons in a SL to perform these
Bloch oscillations before being
scattered\cite{Feldman,Waschke,Dekorsy,Lyssenko}.

If the electric potential drop across $n$ unit cells corresponds
to an optical phonon energy, Bryksin and Firsov\cite{Bryksin1}
predicted another high-field effect: so-called electron-phonon
resonances, where an LO phonon mediates a transition between
localized Wannier-Stark states whose energy separation equals the
phonon energy. These resonances manifest as peaks in the
current-voltage relationship and produce a nonmonotonic
current-voltage dependence. Experimental $I-V$ characteristics of
cubic ZnS films have been identified with such
resonances\cite{Maekawa}. Growing interest in the transport
properties of semiconductor superlattices, e.g. GaAs/AlAs and
InAs/GaSb, has resulted in several recent theoretical studies of
these electron-phonon resonances in
SLs\cite{Bryksin2,Bryksin4,Bryksin3,Rott1,Emin,Wacker2}.
Unfortunately, to our knowledge, electron-phonon resonances have
not been observed experimentally in SL transport within the NDC
regime.

Several choices are available to calculate SL transport in the
high and low field regimes. Wacker and Jauho\cite{Wacker1}
established the relationship in some cases between different
transport calculation methods, including miniband
conduction\cite{Esaki}, Wannier-Stark hopping\cite{Tsu},
sequential tunnelling \cite{Miller}, and the non-equilibrium
Green's function (NGF) formalism\cite{Kadanoff} and their
respective ranges of validity, and reached the conclusion that the
NGF contains the other three as limiting cases. For example, the
semiclassical Boltzmann transport equation (BTE), which can be
solved to any desired degree of numerical accuracy by Monte Carlo
simulations\cite{Jacoboni}, cannot accurately describe transport
under high-field ($eFL\sim\hbar \omega_0$)
conditions\cite{Rott2,Wacker2}.  Use of the generalized
Kadanoff-Baym(GKB) ansatz\cite{Kadanoff,Lipavsky,Spicka} is
limited in validity to slowly varying fields and time dependence.

Rott, Linder and D\"{o}hler\cite{Rott1} employed the Wannier-Stark
hopping transport technique to study the temperature dependence of
electron-phonon resonances. Their method is only applicable to the
high-field regime. Emin and Hart\cite{Emin} obtained a
$\delta-$function electron-phonon resonance in the limit of a very
wide energy band $\Delta\rightarrow\infty$. Bryksin and Kleinert
focused on narrow band($\Delta\rightarrow 0$) transport in a
one-dimensional NGF model with\cite{Bryksin4} and
without\cite{Bryksin3} the GKB ansatz. The approximations made in
Ref.~[\onlinecite{Bryksin3}] included relying on the periodic
nature of the Green's functions in a large homogeneous field and
discarding the Fourier components of the Dyson equations beyond
lowest order. The impurities were modeled with $\delta$-function
potentials (and hence only depended on the product of the impurity
density and the scattering potential\cite{Wacker1,Bryksin3}), and
the SL bandwidths considered were much smaller than optical phonon
energies, in order to obtain analytic expressions for the
field-dependent DOS and the transport.

In our treatment here we use the Kadanoff-Baym-Keldysh NGF
technique\cite{Haug,Kadanoff} to calculate the non-linear quantum
transport. We solve for the field-dependent Green's functions
equations numerically, non-perturbatively including the elastic
impurity scattering by directly inverting the NGF Dyson equations.
These Green's functions, and the resulting density of states
(DOS), depend on the impurity size, strength, and density
separately.  We then calculate the effect of inelastic (phonon)
scattering as a perturbation. Specifically we leave the retarded
Green's function obtained for elastic scattering unchanged, but
alter the less-than Green's function.   A related model, where
impurity scattering was treated exactly within the resonant level
model, and phonon scattering was treated perturbatively,  has been
introduced before\cite{Lipavsky}. In Ref.~[\onlinecite{Lipavsky}]
the analysis focused on comparison of the model results with those
obtained from the GKB ansatz. In our treatment we consider more
general impurity types, and use the solvability of the model to
explore the influence of impurity nature on the electron-phonon
resonances. The resulting $I-V$ curves illuminate the nature of
the electron-phonon resonances in a regime where elastic
scattering dominates.

Because we treat more of the problem numerically we are able to
consider all significant Fourier components of the Dyson equation,
general SL bandwidths and impurity potential size and strength. In
this way we can explore the properties of these electron-phonon
resonances for a wider range of parameters with confidence. Our
approach therefore possesses several advantages over earlier ones:
(i) it employs a more realistic form for impurity potentials that
leads to better approximations for the DOS; (ii) other common
approximations are relaxed, including restrictions on the growth
axis miniband width; (iii)  it is applicable to both high and low
field  transport; and (iv) it does not use the generalized
Kadanoff-Baym ansatz\cite{Kadanoff} to solve the Dyson equation.

The outline of the paper is as follows. In
Sec.~\ref{theory-method}, we show the detailed theoretical and
numerical methods used to calculate the retarded and lesser
Green's functions, from which are extected the DOS and carrier
distribution functions. In Sec.~\ref{results}, the field
dependence of the calculated drift velocity is discussed in
detail. We focus on the electron-phonon resonance peaks
and the influence of impurity scattering and establish a relationship
between earlier results and ours. Our results are
discussed and summarized in Sec.~\ref{discussion}.

\section{Theory and numerical method}
\label{theory-method}

\subsection{High-field Green's function}

In order to study high-field effects, we must use an approach that
is non-perturbative in the electric field. We consider the scalar
potential gauge $\phi ({\bf r},t)=-\bf r\cdot F$, where \textbf{F}
is the time-independent uniform electric field. In the absence of
scattering, the Dyson equation of the retarded Green's function in
the \textbf{k} representation can be written as\cite{Haug}
\begin{eqnarray}
\left[i\hbar \frac{\partial}{\partial t}-\varepsilon ({\bf
k})+ie{\bf F}\cdot \nabla _{\bf k}\right]G_{\phi}^r ({\bf
k},t,{\bf k}',t')=(2\pi)^3\delta ({\bf k-k'})\delta (t-t').\label{dyson-orig}
\end{eqnarray}
We make use of a 14-band superlattice ${\bf K}\cdot{\bf p}$
calculation\cite{fourteenband}  to obtain the superlattice energy
dispersion relations $\varepsilon ({\bf k})$.   Only the lowest
conduction band with growth-axis width $\Delta$ is considered in
the present transport calculations. We assume the electric field
is along the superlattice growth direction (noted as $\|$, and the
in-plane directions are noted as $\perp$). The ${\bf K}\cdot{\bf
p}$ results are then fit to a dispersion relation of the form
\begin{eqnarray}
\varepsilon ({\bf k})=\frac{\hbar^{2}{\bf k}_{\perp}^{2}}{2m^*}
+\frac{\Delta}{2} \left[ 1-\cos(k_{\|}L)\right],
\end{eqnarray}
where $m^*$ is the electron effective mass along the in-plane
directions. Solving Eq.~(\ref{dyson-orig}), we get the retarded
Green's function with no scattering
\begin{eqnarray}
G_{\phi}^r \left( k_{\|},k_{\|}',{\bf k}_{\bot},{\bf
k}_{\bot}',t-t'\right)=-i(2\pi)^3\delta \left[ \hbar
(k_\|-k_\|')-eF(t-t') \right] \delta ({\bf k}_\bot -{\bf k}_\bot
') \theta (t-t')
\nonumber\\ \times
e^{-i\left\{ \left(
\frac{\hbar {\bf k}_\perp ^2}{2m^*}+\frac{\Delta}{2\hbar}\right)
(t-t')+\frac{\Delta}{2eFL}\left[\sin\left(
k_\|'L\right)-\sin\left(k_\|L\right)\right]\right\}}.
\end{eqnarray}
After taking the Fourier transformation ($t-t'\rightarrow \omega$),
\begin{eqnarray}
G_{\phi}^r \left( k_{\|},k_{\|}',{\bf k}_{\bot},{\bf
k}_{\bot}',\omega \right)&=&-\frac{i(2\pi)^3}{eF}\delta({\bf
k}_\perp-{\bf k}_\perp ') \theta(k_\|-k_\|')
\nonumber\\ &&\times
e^{-\frac{i}{eF} \left\{ \left(-\hbar\omega+\frac{\hbar ^2 {\bf
k}_\perp^2}{2m^*} +\frac{\Delta}{2}\right)\left(k_\| -k_\|'\right)
+\frac{\Delta}{2L} \left[\sin(k_\|'L)-\sin(k_\|L)\right]
\right\}}.
\end{eqnarray}
The corresponding gauge invariant spectral function is
\begin{eqnarray}
A({\bf k},\omega) &=& i\left[ G^r({\bf k},\omega)-G^a({\bf
k},\omega)\right]\nonumber \\
&=& 2\pi \sum_{n=-\infty}^{\infty}
J_n\left[-\frac{\Delta}{eFL}\cos(k_\|L)\right] \delta
\left(\hbar\omega-\frac{\hbar^2{\bf k}_\bot^2}{2m^*}-
\frac{\Delta}{2}-\frac{neFL}{2}\right).\label{spectral-func}
\end{eqnarray}
The spectral function satisfies the sum rule $(\hbar /2\pi)\int d\omega A({\bf
k},\omega)=1$. The density of states is
\begin{eqnarray}
\rho({\bf k_\perp},\omega)&=&\int^{\pi/L}_{-\pi/L}
\frac{dk_\|}{2\pi}A( {\bf k},\omega)\nonumber\\ &=&\frac{2\pi}{L}
\sum^{\infty}_{n=-\infty}J^2_n\left(-\frac{\Delta}{2eFL}\right)
\delta\left(\hbar\omega-\frac{\hbar^2{\bf k^2_\perp}}{2m^*}-
\frac{\Delta}{2}-neFL\right).
\end{eqnarray}
The energy of the Wannier-Stark level $n$ superlattice cells away
is $neFL$. A level with index $n$ can contribute to the DOS if the
miniband width $\Delta$ is big enough, and its contribution is
proportional to $J^2_n\left(-\frac{\Delta}{2eFL}\right)$. This
result coincides with Wacker's results based on Wannier-Stark
bands\cite{Wacker1}. In the NGF approach of Bryksin\cite{Bryksin3}
the terms with $n>0$ vanish because the approximation
$\Delta\rightarrow 0$ is used.

\subsection{Elastic impurity scattering and $G^r$}\label{elastic-gr}

For a single impurity at ${\bf R}$, the Dyson equation for the
retarded Green's function is
\begin{eqnarray}
G_{\bf R}^r({\bf k,k}',\omega) &=& G_\phi^r({\bf k,k}',\omega)+
\int \frac{d^3{\bf q_1}}{(2\pi)^3}\int\frac{d^3{\bf
q_2}}{(2\pi)^3} G_\phi^r({\bf k,q}_1,\omega)
\nonumber\\ && \times
e^{-i(\bf q_1-q_2)\cdot {\bf R}} V({\bf q_1,q_2},\omega)G_{\bf
R}^r({\bf q_2,k'},\omega).
\end{eqnarray}
For elastic impurity scattering, we have $V({\bf
q_1,q_2},\omega)=V({\bf q_1-q_2})$. In our calculations, we
consider a one-dimension system and use a Gaussian impurity
potential with range $a$ along the growth direction
\begin{eqnarray}
V({\bf r_1-r_2})&=&V_0 e^{-{\left(r_{1\|}-r_{2\|}
\right)^2}/{a^2}}.
\end{eqnarray}
The Fourier transform of the impurity potential is
\begin{eqnarray}
V({\bf q_1-q_2})&=&V_0(2\pi)^2a\sqrt{\pi}\delta({\bf
q_{1\perp}-q_{2\perp}})e^{-{a^2\left(q_{1\|} -q_{2\|}
\right)^2}/{4}} .
\end{eqnarray}
Omitting the perpendicular component ${\bf k_\perp,q_\perp}$, we
find for the Dyson equation
\begin{eqnarray}
G^r_R\left(k_\|,k_\|',\omega\right)&=&G^r_\phi\left(k_\|,k_\|',
\omega\right)+\int \frac{dq_\|}{2\pi}\int\frac{dq_\|'}{2\pi}
G^r_\phi\left(k_\|,q_\|,\omega\right)
\nonumber\\&&\times
e^{-i\left(q_\|-q_\|'\right)
R}V(q_\|-q_\|')G^r_R\left(q_\|',k_\|',\omega\right).
\end{eqnarray}
The above Green's function is for only one impurity at site $R$
and for a single scattering event. When all scattering events at
that site are included, the Green's function and corresponding
T-matrix are
\begin{eqnarray}
G^r_R\left(k_\|,k_\|',\omega\right)=G^r_\phi\left(k_\|,k_\|',\omega\right)
+\int\frac{dq_\|}{2\pi}\int\frac{dq_\|'}{2\pi}G^r_\phi\left(k_\|,
q_\|,\omega\right)
\nonumber\\ \times
 T'_R\left(q_\|,q_\|',\omega\right)G^r_R\left(
q_\|',k_\|',\omega\right),\label{retarded-gf1}
\end{eqnarray}
\begin{eqnarray}
T'_R\left(q_\|,q_\|',\omega\right)=e^{-i\left(q_\|-q_\|'\right)R}
V\left(q_\|-q_\|'\right)+\int\frac{dp_\|}{2\pi}\int\frac{dp_\|'}{2\pi}
V\left(q_\|-p_\|\right)
 \nonumber\\\times
e^{-i\left(q_\|-p_\|\right)R}G^r_\phi\left(p_\|,
p_\|',\omega\right)T'_R\left(p_\|',q_\|',\omega\right).\label{t-matrix1}
\end{eqnarray}
We assume the impurity density is sufficiently dilute that we can
approximate the self energy by its impurity-averaged form.  The
self energy is then
\begin{eqnarray}
\Sigma^r\left(q_\|,q_\|',\omega\right)=c\int
dRT_R\left(q_\|,q_\|', \omega\right),\label{self-energy}
\end{eqnarray}
where $c$ is the {\it line} concentration of impurities. $T_R$
differs from $T'_R$ of Eq.(\ref{t-matrix1}); we apply the
self-consistent Born approximation\cite{Lake,Haug}  and we replace
$G^r_\phi$(no scattering) in Eq.(\ref{retarded-gf1})
by $G^r$(including scattering) to obtain $T_R$. We find
\begin{eqnarray}
T_R\left(q_\|,q_\|',\omega\right)=e^{-i\left(q_\|-q_\|'\right)R}
V\left(q_\|-q_\|'\right)+\int\frac{dp_\|}{2\pi}\int\frac{dp_\|'}{2\pi}
V\left(q_\|-p_\|\right)
\nonumber\\\times
e^{-i\left(q_\|-p_\|\right)R}G^r\left(p_\|,
p_\|',\omega\right)T_R\left(p_\|',q_\|',\omega\right).\label{t-matrix2}
\end{eqnarray}
The final Dyson equation for the retarded Green's function $G^r$ is thus
\begin{eqnarray}
G^r\left(k_\|,k_\|',\omega\right)=G^r_\phi\left(k_\|,k_\|',\omega\right)
+\int\frac{dq_\|}{2\pi}\int\frac{dq_\|'}{2\pi}G^r_\phi\left(k_\|,
q_\|,\omega\right)
\nonumber\\ \times
\Sigma^r\left(q_\|,q_\|',\omega\right)G^r\left(
q_\|',k_\|',\omega\right).\label{retarded-gf2}
\end{eqnarray}

Similar equations for the resonant level model (RLM) were obtained
by Jauho\cite{Haug}. Here our approach is rather different.
Instead of exploring the relationship of these equations to the
Boltzmann equation\cite{Haug} we plan to apply the equation
numerically to investigate in detail electron-phonon resonances.

Eqs. (\ref{t-matrix2}) and (\ref{retarded-gf2}) form a closed set
of integral equations. In our numerical method, we use a uniformly
distributed grid for k-space and R-space integration, hence the
double integral becomes a product of matrices. The sizes of these
matrices $G^r$ and $T_R$ are typically $600\times 600$ (15
Brillouin zones and 40 points per zone). The R-space integration
in Eq. (\ref{self-energy}) involves about 500 points with a
$1000\rm{\AA}$ grid spacing. To solve the equation set, we first
use $G^r_\phi$ as an initial guess for $G^r$ and perform the
iteration to obtain a self-consistent solution. Sometimes the
iterative process does not converge, in which case we adjust the
initial retarded Green's function, matrix size, or (k,R)-space
integral mesh.

\subsection{Polar optical phonon scattering and $G^<$}

We assume phonon scattering is very weak compare to elastic
scattering. In this limit the Wannier-Stark level broadening is
mainly governed by elastic impurity scattering --- thus the
densities of states are determined by the elastic impurity
scattering. We obtain the retarded Green's function, which
determines the DOS, by the procedure of Sec.~\ref{elastic-gr}
(neglecting phonon scattering). The approach to $G^<$ must be
different, however, for this function describes the carrier
dynamics. When only elastic impurity scattering is considered,
$G^<=-G^>=-1/2[G^r-G^a]=iA/2$, and the net current vanishes. In
order to obtain a non-zero current,  phonon scattering must be
taken into account in $G^<$.

We evaluate the  lesser Green's function directly from the Keldysh
equation\cite{Kadanoff}, but with a perturbative approach.
\begin{eqnarray}
G^<=(1+G^r\Sigma^r)G_0^<(1+\Sigma^aG^a)+G^r\Sigma^<G^a,\label{lesser-gf1}
\end{eqnarray}
where the double momentum labels for each of the Green's functions
and self-energies are implicit and all the products in the
equation imply integration over the adjacent momentum label. Here
$G_0^<$ is the lesser Green's function without scattering and is
equal to $-[G^r_\phi-G^a_\phi]/2 = iA({\bf k},\omega)/2$. The
spectral function neglecting scattering, $A({\bf k},\omega)$
(Eq.(\ref{spectral-func})), has several poles of the form
$\delta(E-E_n)$. Scattering makes the poles move away from these
energies $E_n$; the first term in Eq.~(\ref{lesser-gf1}) is
proportional to $(G^r_0)^{-1}G^<_0(G^a_0)^{-1}\propto
(E-E_n)^2\delta(E-E_n)$ near the former singular points. As a
consequence, the first term in Eq.~(\ref{lesser-gf1}) will vanish,
so Eq.~(\ref{lesser-gf1}) simplifies to
\begin{eqnarray}
G^<=G^r\Sigma^<G^a.\label{lesser-gf2}
\end{eqnarray}
Here the self energy is $\Sigma^<=\Sigma^<_{imp}+\Sigma^<_{pho}$.
Based on analytic continuation\cite{Haug}, the self energy for
impurity scattering is
\begin{eqnarray}
\Sigma^<_{imp}=c\int dR T^r_RG^<T^a_R.\label{lesser-gf-imp}
\end{eqnarray}
For polar optical phonon scattering, the corresponding self energy
is expressed as
\begin{eqnarray}
\Sigma^<_{pho}=\int\frac{dq}{2\pi}M^2_q\left[n_BG^<\left(\omega-
\omega_q\right)+\left(n_B+1\right)G^<\left(\omega+\omega_q\right)
\right],\label{lesser-gf-pho}
\end{eqnarray}
where $M_q$ is the electron-phonon coupling matrix element, here
treated as a constant, and
$n_B=1/\left\{\exp[\hbar\omega_q/(k_BT)]-1\right\}$ is the
Bose-Einstein occupation factor for LO phonons. The frequency of
the LO phonons $\omega_q$ is also treated as a constant. The
temperature we use is 77K. Earlier work\cite{Rott1,Wacker2} shows
higher temperatures will strongly depress the electron-phonon
resonances.

The following symmetry holds when there is only elastic
scattering: $G(k,k',\omega+\omega_0)=e^{i(k-k')\frac{\hbar
\omega_0}{eF}}G(k,k',\omega)$ or alternately $G(k,k',t,t') =G(k,
t-t')\delta[k-k'-\frac{eF}{\hbar}(t-t')]$. We assume this symmetry
is still present when  phonon scattering is included as a
perturbation. As a first order approximation, we replace $G^<$ in
Eqs. (\ref{lesser-gf-imp})and (\ref{lesser-gf-pho}) with
$\frac{i}{2}A_{imp}=-\frac{1}{2}[G^r_{imp}-G^a_{imp}]$. The lesser
Green's function of Eq.(\ref{lesser-gf2}) then becomes
\begin{eqnarray}
G^< &=& G^r_{imp}\left[\Sigma^<_{imp}+\Sigma^<_{pho}\right]
G^a_{imp}\nonumber\\ &=& G^r_{imp}\left\{c\int dR T^r_R
\frac{i}{2}A_{imp}T^a_R\right.
\nonumber\\&&
\left.
+\int\frac{dq}{2\pi}M^2_q\left[n_B
\frac{i}{2}A_{imp}(\omega-\omega_q)+(n_B+1)\frac{i}{2}A_{imp}
(\omega+\omega_q)\right]\right\}G^a_{imp}\nonumber\\ &=&
\frac{i}{2}A_{imp}+G^r_{imp}\left\{\int\frac{dq}{2\pi}M^2_q
\left[\frac{i}{2}n_Be^{-i(k-k')\frac{\hbar\omega_q}{eF}}
A_{imp}(\omega)\right.\right.
\nonumber\\&&
\left.\left.
+\frac{i}{2}(n_B+1)e^{i(k-k') \frac{\hbar
\omega_q}{eF}}A_{imp}(\omega)\right]\right\}G^a_{imp}.\label{lesser-gf3}
\end{eqnarray}
The first term has been simplified because $iA_{imp}$ is a
solution for the equation $G^<_{imp}=G^r_{imp}\left[c\int dR T^r_R
G^<_{imp}T^a_R\right]G^a_{imp}$. The variables on the right side
of Eq. (\ref{lesser-gf3}) are all known, so $G^<(k,k',\omega)$ can
be found. The current density $J$ depends on $G^<$ according to
\begin{eqnarray}
J=e\int^{\pi/L}_{-\pi/L}\frac{dk}{2\pi}\left[(-i)G^<(k,t=0)\frac{1}{\hbar}
\frac{\partial \varepsilon(k)}{\partial k}\right].
\end{eqnarray}

\section{Results}
\label{results}

\subsection{Impurity scattering effects on electron-phonon resonance}

We expect the properties of the electron-phonon resonances to
depend significantly on the relative magnitudes of the growth axis
bandwidth of the lowest conduction miniband and the optical phonon
energy. If the bandwidth $\Delta$ is very large compared to the
optical phonon energy, then the electron-phonon resonances can be
well described within a Wannier-Stark hopping picture, whereas in
the other limit the electron-phonon resonances are better
described within a sequential tunnelling picture. We present here
detailed results for the drift velocity as a function of electric
field for two superlattices, one in each regime.

The first superlattice consists of 12 monolayers of GaAs and 6
monolayers of AlAs (unit cell length 50.9\AA).  The resulting
width of the lowest miniband is 20.3meV. We use the value for the
GaAs optical phonon energy ($\hbar\omega=36$meV) for this
GaAs/AlAs SL, so the ratio of the bandwidth to the optical phonon
energy $\Delta/\hbar\omega =0.56$. The other structure is 7
monolayers of InAs and 12 monolayers of GaSb (unit cell length
57.8\AA), which has a very broad lowest conduction miniband width
of 150meV. The optical phonon energy of InAs ($\hbar\omega=30$meV)
is used for the InAs/GaSb superlattice, and hence
$\Delta/\hbar\omega = 5$. In both cases, the zone center gap
between the lowest and second-lowest minibands is larger than
350meV. Thus, the occupation of higher minibands is very small and
we can neglect conduction from higher minibands and interminiband
Zener tunneling.  For convenience in our comparison we approximate
the period for the two types of superlattices with the same value,
$L=57$\AA.

We have numerically calculated the drifty velocity for fields near
the electron-phonon resonance value $neFL=\hbar\omega$, which
typically ranges from $0.1\times 10^7$ V/m to $1.0\times 10^7$
V/m. Because we treat phonon scattering as a perturbation, the
calculated drift velocity is directly proportional to the
electron-phonon coupling coefficient $M^2_q$. We choose the
coupling constant to make the drift velocity realistic ($\sim
10^3$m/s). To be consistent, the calculated effective relaxation
time $\tau_{eff}$ from phonon scattering should be longer than the
relaxation time from impurity scattering: $\tau_{eff}>10$ps.

\begin{figure}
\includegraphics[width=5in]{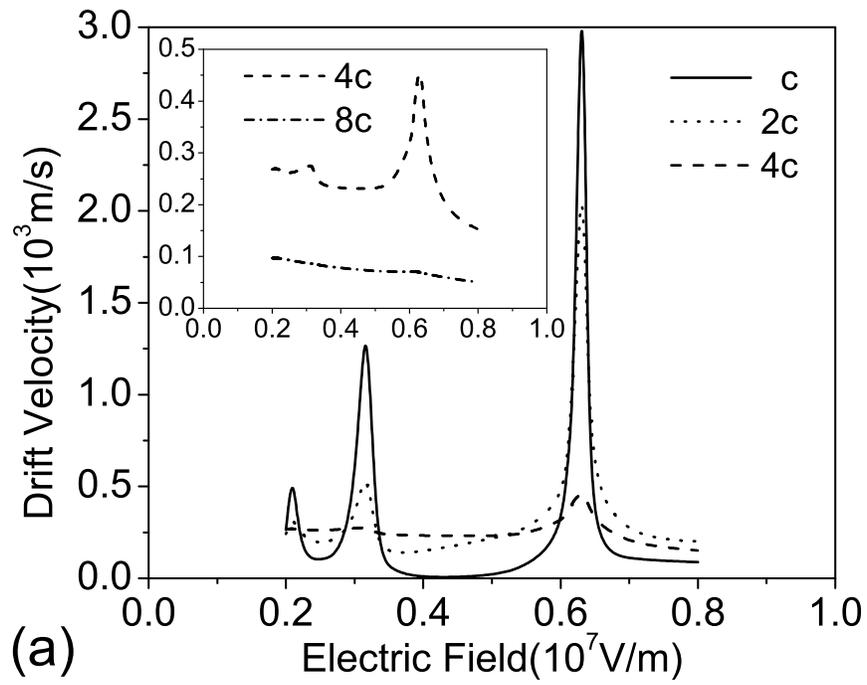}
\includegraphics[width=5in]{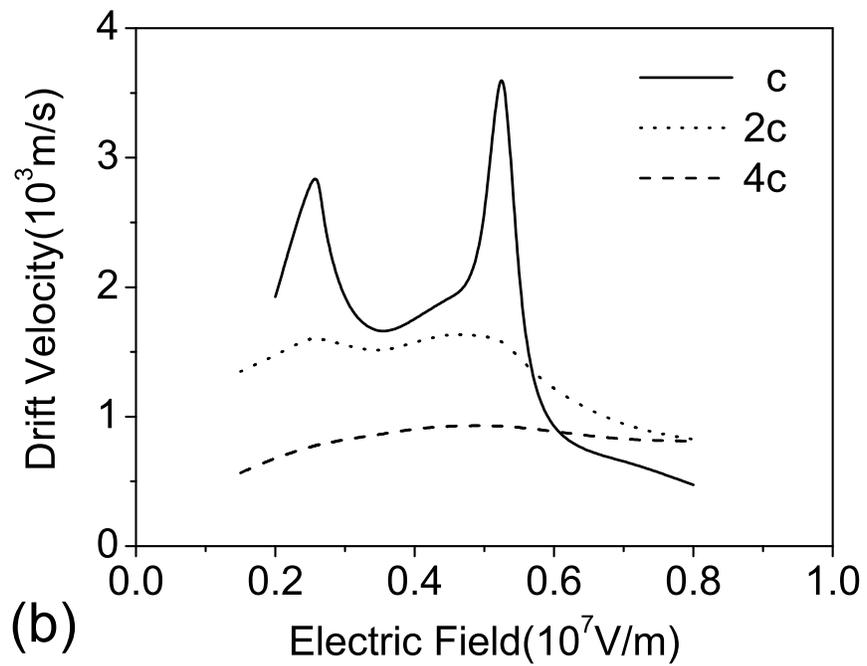}
\caption{Shaoxin Feng {\it et al.} \label{dep-imp-conc}}
\end{figure}

In Fig. \ref{dep-imp-conc}(a), we show how the density of
impurities affects the electron-phonon resonance peaks in the
narrow bandwith superlattice, and in Fig. \ref{dep-imp-conc}(b)
results for the broad bandwidth one. The line concentration of
impurities ranges from $1/(464\rm\AA)$ (defined as $c$ in
Fig.~\ref{dep-imp-conc}) to $1/(58\rm\AA)$. These values
correspond to $1.0\times 10^{16} \rm cm^{-3}$ and $5.1\times
10^{18} \rm cm^{-3}$, respectively,  in 3-dimensions. We see that
large impurity densities will produce weaker electron-phonon
resonance peaks. When the impurity density is $4c$, there is only
one peak in the velocity-field curve for the broad bandwidth SL
(Fig. \ref{dep-imp-conc}(b)). Indeed, the position of the peak is
not at a resonance, so it's not an electron-phonon resonance at
all. Instead, it's the maximum of the entire velocity-field curve,
which is the starting point of NDC. In Fig.~\ref{dep-imp-conc}(a),
all the electron-phonon resonance peaks are located in the NDC
region. In Fig. \ref{dep-imp-conc}(b), when the impurity
scattering is very strong the electric field corresponding to the
maximum current enters the region of our interest. This peak
flattens the electron-phonon resonance peaks. The calculated DOS,
instead of exhibiting Wannier-Stark levels, is closer to a
zero-field DOS.

\begin{figure}
\includegraphics[width=5in]{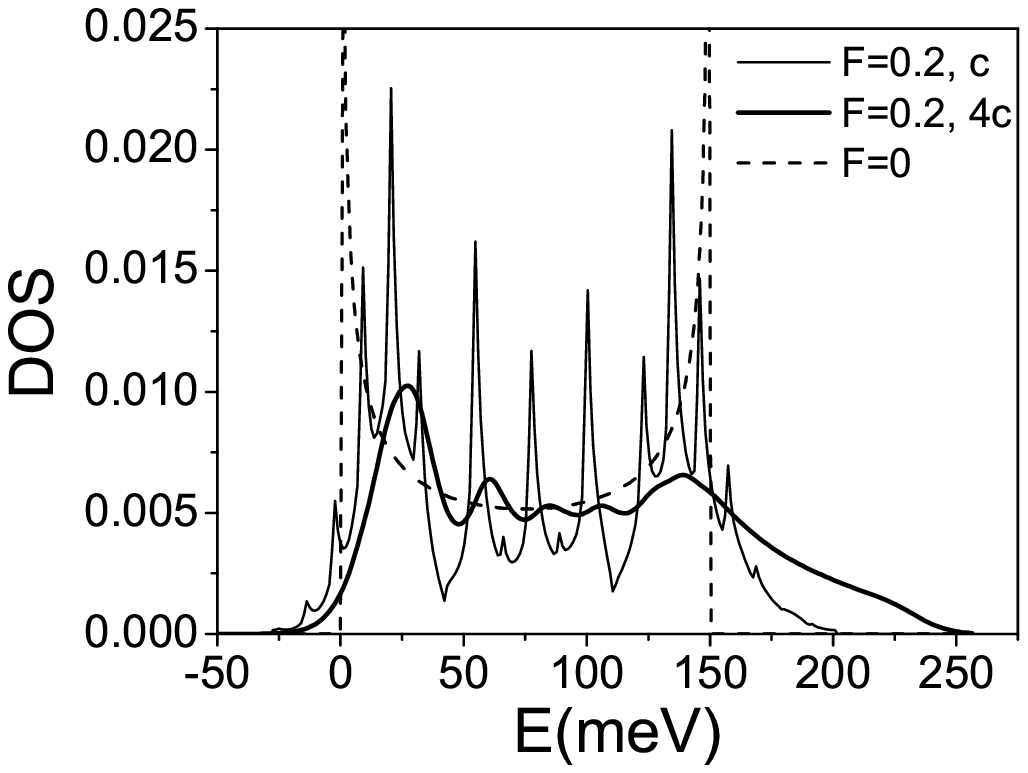}
\caption{Shaoxin Feng {\it et al.} \label{dos-0field}}
\end{figure}

In Fig.~\ref{dos-0field}, the solid lines show the DOS for
different impurity concentrations under the same electric field
$F=0.2\times10^7$V/m in broad-band superlattice. The DOS for a low
impurity density ($c$=1/(464\AA), thin line in the
Fig.~\ref{dos-0field}) clearly shows the Wannier-Stark levels. But
the DOS for a higher impurity concentration ($c$=1/(116\AA), thick
line) tends to be very similar to the zero-field result (which is
shown as the dashed line). Properly approaching the zero-field DOS
and the appearance of the maximum of current in the velocity-field
curves indicate our method is valid for low-field situations. This
contrasts with the Wannier-Stark hopping method that is often used
in high-field transport, which will diverge under low-field
conditions\cite{Wacker1,Rott1}.

\begin{figure}
\includegraphics[width=5in]{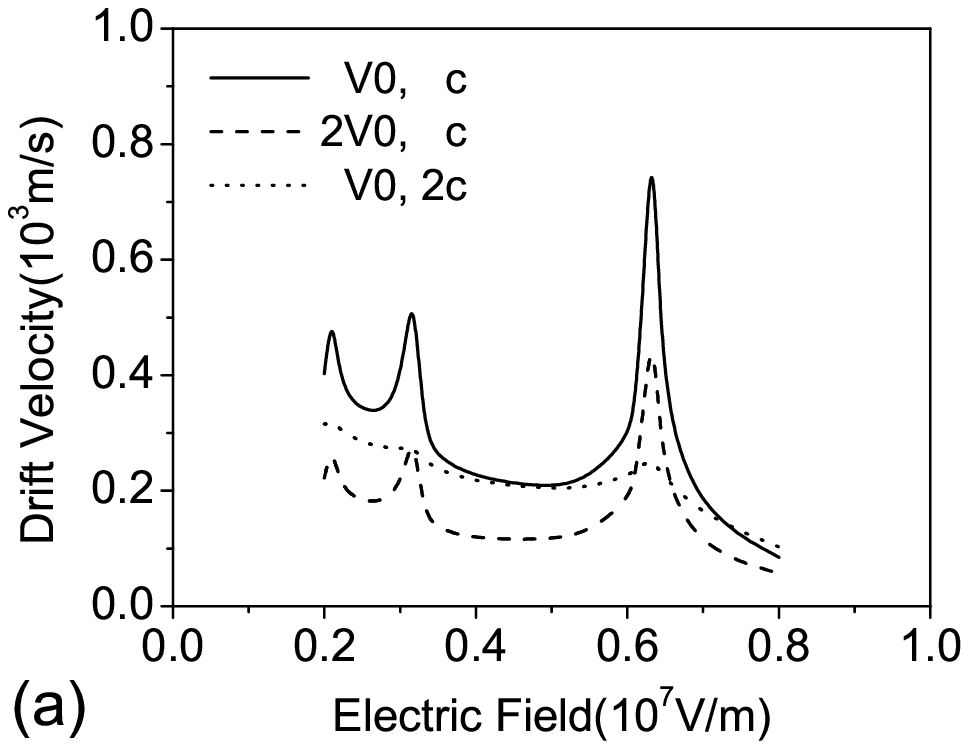}
\includegraphics[width=5in]{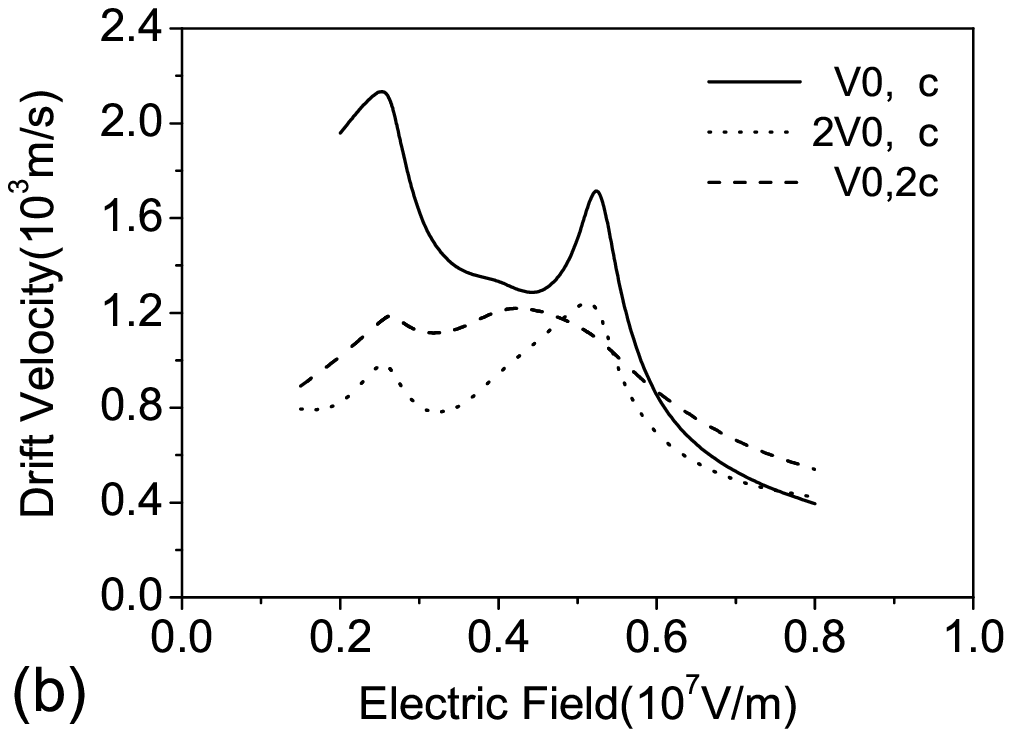}
\caption{Shaoxin Feng {\it et al.} \label{imp-stre-conc}}
\end{figure}

\begin{figure}
\includegraphics[width=5in]{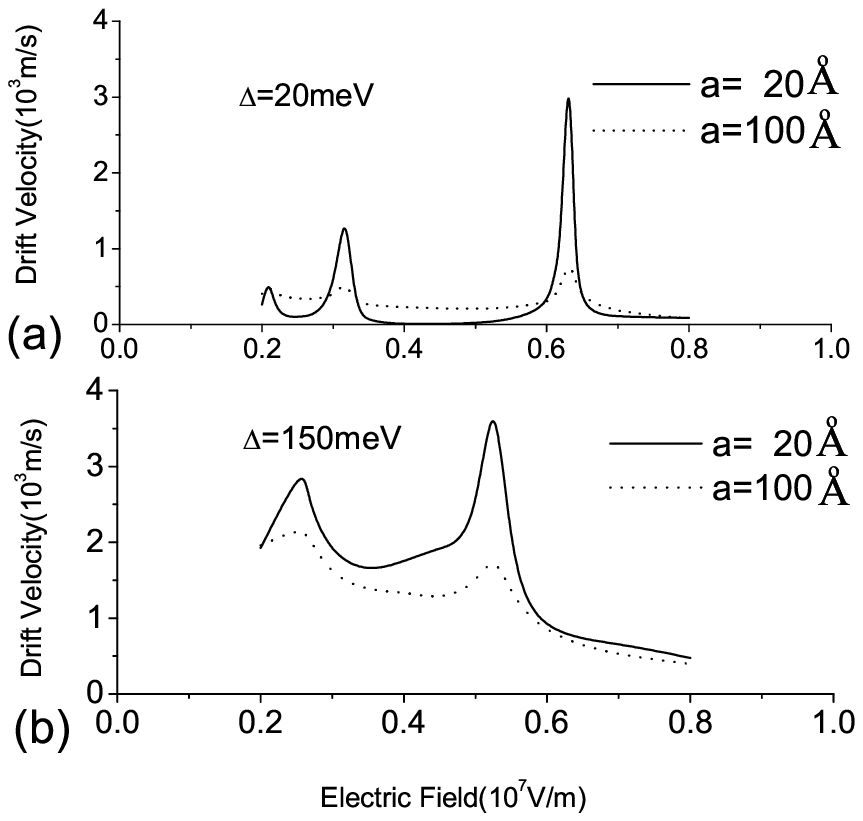}
\caption{Shaoxin Feng {\it et al.} \label{imp-size}}
\end{figure}

In Fig. \ref{imp-stre-conc}, we compare the effects of the
impurity strength $V_0$ and impurity density $c$ on the
electron-phonon resonances. We find that the impurity strength can
diminish the drift velocity considerably, but does not affect the
height of the electron-phonon resonances as much as the
concentration of impurities does. In Fig. \ref{imp-size}, we
change the impurity size but keep $V_0a$ fixed to observe changes
in the electron-phonon resonances. When the impurity size
increases, the resonance becomes weaker.

\begin{figure}
\includegraphics[width=5in]{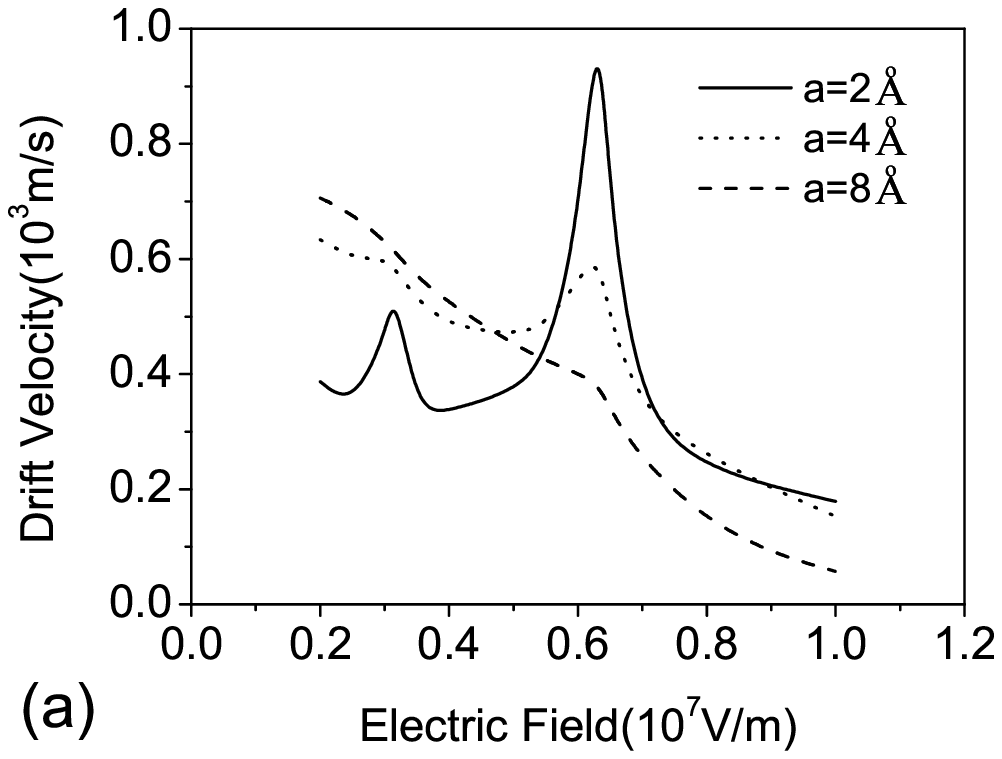}
\includegraphics[width=5in]{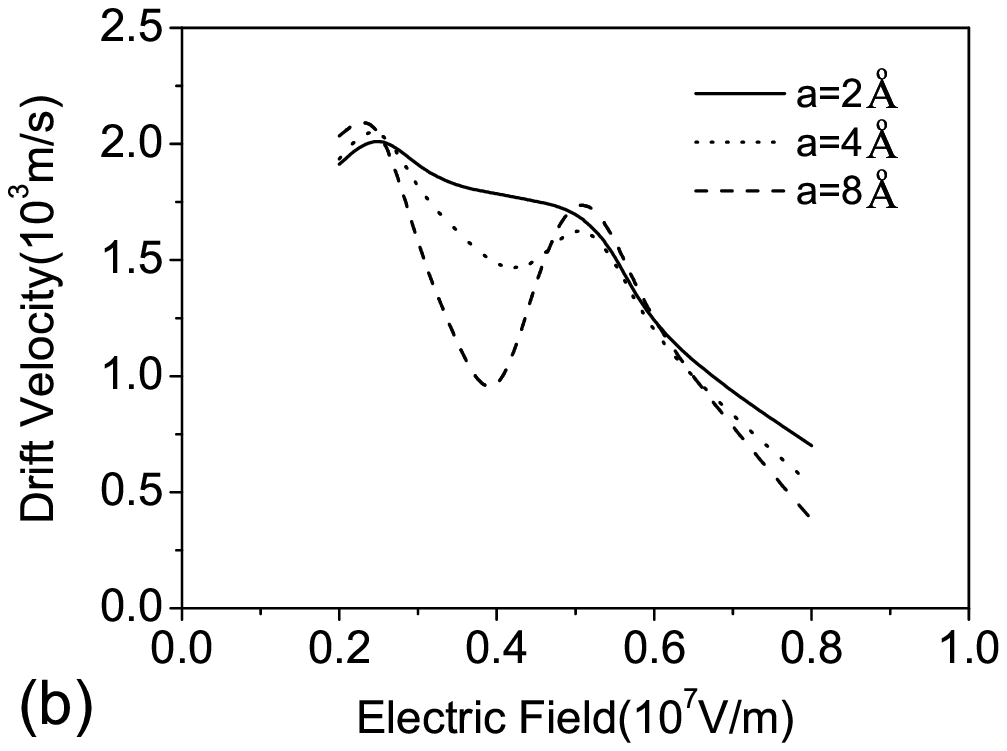}
\caption{Shaoxin Feng {\it et al.} \label{dirty-imp-size}}
\end{figure}

Figs. \ref{imp-stre-conc} and \ref{imp-size} are calculated
results for relatively perfect materials. We also model more
imperfect materials by employing an impurity density of
$1/(19\rm\AA)$ (corresponding to $1.46\times 10^{20}\rm cm^{-3}$
in  three dimensions), and use a rather small value for the
impurity strength $V_0$. The impurity size produces different
effects on the electron-phonon resonance peaks for different
bandwidths. In Fig.~\ref{dirty-imp-size}(a), for the narrow
bandwidth SL, impurities of larger size (with $V_0a$ constant)
flatten the electron-phonon resonance peaks. This contrasts with
Fig. \ref{dirty-imp-size}(b), for the broad bandwidth SL, where
larger impurities {\it enhance} the electron-phonon resonances
stronger. This phenomena can be explained by comparing with the
one-dimensional square potential barrier scattering problem. The
reflection coefficient of a scattering electron is not a monotonic
function of the barrier size (still keeping $V_0a$
constant)\cite{Schiff}. Instead, there is a maximum value for the
reflection probability. We take $\Delta/2$ as the average energy
of incoming electrons and calculate the impurity size that
maximizes the reflection coefficient. For the narrow bandwidth SL,
$a_{max}=11.3\rm\AA$, whereas for the broad bandwidth SL,
$a_{max}=1.5\rm\AA$. Hence impurity scattering generates different
trends for narrow and broad bandwidth SLs in the region
$2\rm{\AA}<a<8\rm{\AA}$.

There is another feature distinguishing relatively clean and dirty
superlattices. In the clean case the electron-phonon resonance in
the narrow bandwidth SL is stronger than in broad bandwidth one.
But  in the dirty case, the electron-phonon resonance in the
narrow bandwidth SL is weaker than in the broad bandwidth one,
e.g. when the impurity size is $a=8\rm\AA$. The reason for this is
as follows: when $V_0a$ is large, all the collisions between
electrons and impurities are essentially total reflection. In this
case, impurity scattering in a broad bandwidth SL is stronger than
in a narrow bandwidth one because the spatial extent of the
Wannier-Stark wave function $\Delta/eF$ is proportional to the
bandwidth $\Delta$. Thus electrons in a broad bandwidth SL
experience more collisions in a Wannier-Stark oscillation period.
But when $V_0a$ is small, the collisions have both substantial
transmission and reflection. The incoming electrons in a broad
bandwidth SL have more energy and a larger transmission
coefficient than in a narrow bandwidth SL. This makes impurity
scattering in broad bandwidth SLs weaker than in narrow bandwidth
ones, which is the opposite of the clean case.

\subsection{Comparison with earlier results}

\begin{figure}
\includegraphics[width=5in]{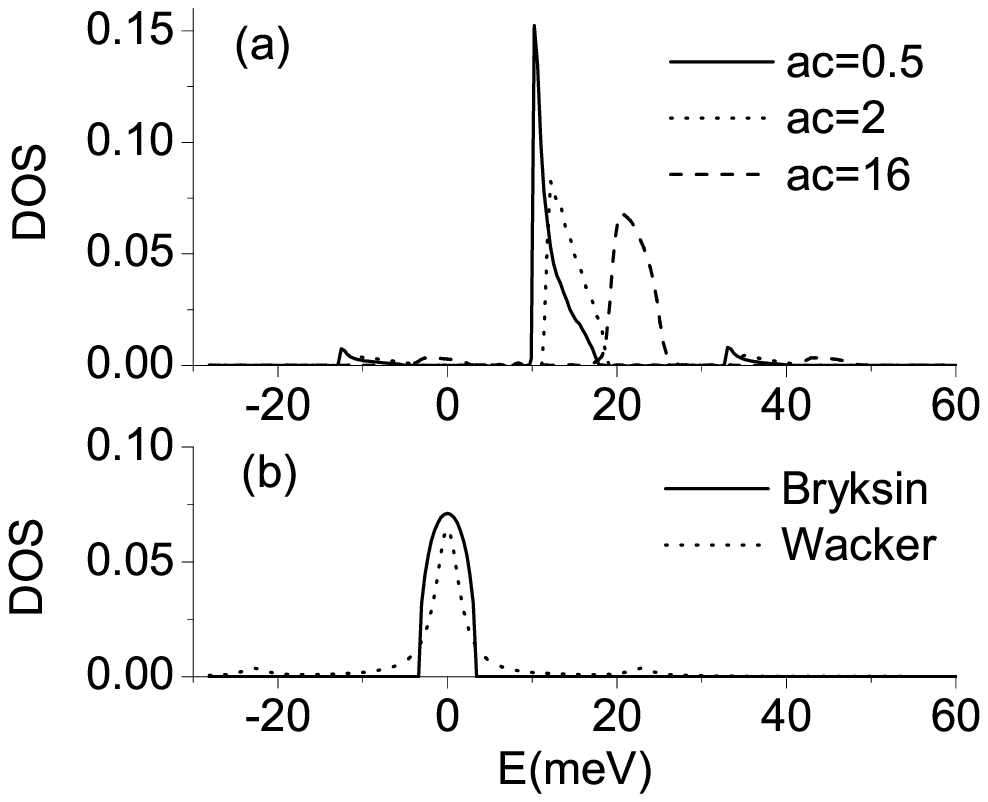}
\caption{Shaoxin Feng {\it et al.} \label{dos-bryk-wack}}
\vspace{10cm}
\end{figure}

We have also reproduced the earlier work of Bryksin and
Kleinert\cite{Bryksin3} and find a relation between Bryksin's
parameter for impurity scattering and ours: $\hbar\sqrt{U}\approx
V_0\sqrt{ac}$. However, this relation has a limited range of
applicability because the DOS of a Wannier-Stark level generally
does not have the half-elliptical shape employed by Bryksin,
\begin{equation}
\propto 1/(2\pi\hbar U)Re\sqrt{4U-\omega^2}.
\end{equation}
Fig.~\ref{dos-bryk-wack} shows the DOS for different parameter
sets and compares them to Bryksin's and
Wacker's\cite{Wacker1,Wacker2} results. When $ac$ is very large,
the DOS tends to have a half-elliptical shape. From Fig.
\ref{dos-bryk-wack}, we also can see a shift of the DOS. If there
is no scattering, the main band is centered at $E=\Delta/2$.
Impurity scattering will shift all bands to higher energies. When
$ac$ is big enough the shape of the DOS is Bryksin-like and the
energy shift has the form $E_{center}=E_{0center}+\delta
E=\Delta/2+ \sqrt{\pi}V_0 ac$. Bryksin's DOS centers on the
zero-energy point because they used $\Delta\rightarrow 0$.
Wacker's DOS also remains centered on the point $E=0$ because they
chose localized Wanner-Stark states as the basis of the wave
function, as have many others\cite{Lake}. Wacker's DOS also uses
the approximation of a constant scattering self-energy,
\begin{eqnarray}
G^r(E,{\bf k_\perp})=\sum_n\frac{J^2_n\left(\Delta/2eFL
\right)}{E-neFL-E_{\bf k_\perp}+(i\Gamma/2)},
\label{wacker-retarded-gf}
\end{eqnarray}
where $-i\Gamma/2$ is the self-energy (in Fig. \ref{dos-bryk-wack}
we adopt $\Gamma=2\hbar\sqrt{U}$). The shape of Wannier-Stark
state using a constant self-energy is not realistic. However, Eq.
(\ref{wacker-retarded-gf}) does produce the appropriate side band.
Bryksin's DOS uses the miniband approximation $\Delta/2eFL\ll 1$
and neglects the side band, and therefore obtains a
field-independent DOS. For the narrow-band SL, $\Delta=$20meV, the
central band occupies $J^2_0(\Delta/2eFL)=90.7\%$ of the entire
density when a typical electric field $F=0.4\times 10^7V/m$ is
used.

\begin{figure}
\includegraphics[width=5in]{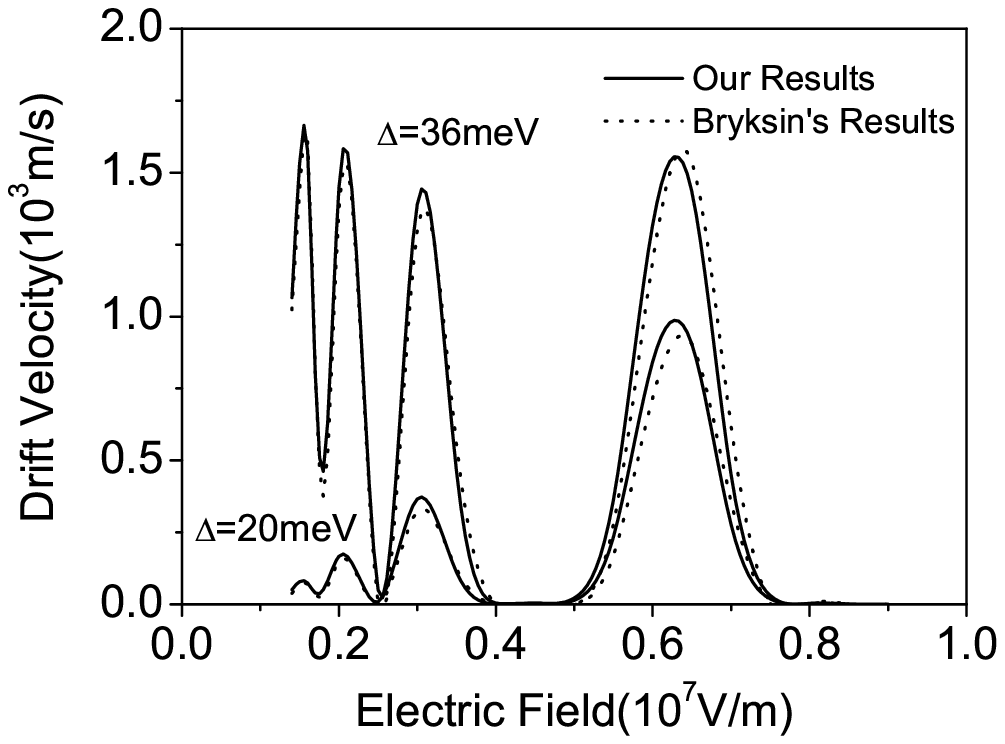}
\caption{Shaoxin Feng {\it et al.} \label{reprod-bryk}}
\end{figure}

In Fig. \ref{reprod-bryk}, we reproduce Bryksin's velocity-field
curve using a very large value for $ac$. The agreement is very
good. When $ac$ is small, the DOS deviates from the
half-elliptical shape and becomes asymmetric. The asymmetry
manifests in a low energy side which has a larger density than the
high energy side [see Fig.~\ref{dos-bryk-wack}(a)]. For a
realistic superlattice, $ac<1$, so the half-elliptical shape will
typically not exist.

We have also calculated the approximate momentum relaxation times
($\tau$) for different impurity parameters and present them in
Table 1. The $\tau$'s are calculated by fitting the retarded
Green's function $|G^r(k,t)|$ to an exponential decay
$e^{-t/\tau}$. In Bryksin's limit (using the half-elliptical DOS),
$G^r(t)$ has the form $J_1(2\sqrt{U}t)/\sqrt{U}t$, which
oscillates in time and exhibits a power-law decay. In our
calculations, the decay of $G^r$ is neither Bryksin-like nor
exponential. The decay of $G^r(t)$ is Bryksin-like when $t$ is
short, and the decay is exponential when $t$ is long. In the
intermediate region, the decay displays a complex shape that
depends on the specific impurity scattering parameters.

\section{Discussion}
\label{discussion}

The advantages of our computational method are: (i) we treat
realistic scattering strengths and obtain a realistic DOS; (ii)
many common approximations are relaxed, such as the small
bandwidth ($\Delta\rightarrow 0$), the high-field
($\Omega\tau_{eff}\gg 1$) and the relaxation time approximations;
(iii) compared to the Wannier-Stark hopping method that was mainly
used under high-field conditions, our method is applicable to both
high and low field  transport; and (iv) we do not use the
generalized Kadanoff-Baym ansatz\cite{Kadanoff} to solve the Dyson
equation.

In our calculations, we used the approximation that phonon
scattering is weak compared to impurity scattering. It is
straightforward in principle to eliminate this approximation in
the non-equilibrium Green's function method by adding self-energy
terms for phonon scattering $\Sigma^r_{pho}$\cite{sigma} to
Eq.(\ref{retarded-gf2}).
\begin{eqnarray}
\Sigma^r_{pho}&=&\int\frac{dq}{2\pi}M^2_q\left[n_BG^r\left(\omega-
\omega_q\right)+\left(n_B+1\right)G^r\left(\omega+\omega_q\right)
\right.\nonumber\\&&\left.+i\int\frac{d\omega'}{2\pi}
G^<(\omega-\omega')\left(\frac{1}{\omega'-
\omega_0+i0^+}-\frac{1}{\omega'+\omega_0+i0^+}\right)\right].
\label{self-energy-pho}
\end{eqnarray}

If we ignore the last term in Eq.(\ref{self-energy-pho}), which
corresponds to a low carrier concentration approximation, then Eq.
(\ref{retarded-gf2}) is a closed equation for the retarded Green's
function. Based on the retarded Green's function, we can use
iteration or some other method to solve the Quantum Kinetic
Equation (\ref{lesser-gf2}) to obtain the lesser Green's function.
If the low carrier concentration approximation is relaxed, Eq.
(\ref{retarded-gf2}) and Eq. (\ref{lesser-gf2}) are coupled and
more iterations are needed. In the both cases, the computational
effort is much greater than the present approach.

\section*{Acknowledgment}
We would like to acknowledge useful discussions with S. J. Allen.
This work was supported by NSF Grant No. ECS0000556.

\newpage
\begin{center}
Table 1. Calculated approximate relaxation times for different
impurity scattering strengths.

\begin{tabular}{|c|c|c|c|c|}\hline
\multicolumn{3}{|c|}{Impurity parameters} & \multicolumn{2}{c|}
{Approximate relaxation time(ps)} \\ \hline a(\AA) & V0(eV) &
c(1/\AA) & $\Delta =20$meV & $\Delta =150$meV \\ \hline 20 & 0.1 &
1/464 & 8.95 & 1.73 \\ \hline 20 & 0.1 & 2/464 &3.09&0.377
\\\hline 20&0.1&4/464&0.905&0.116 \\\hline 20&0.1&8/464&0.470&|
\\\hline 100&0.02&1/464&3.32&1.18 \\\hline 100&0.04&1/464&2.12
&0.683\\\hline 100&0.02&2/464&0.940&0.321\\\hline
\end{tabular}
\end{center}

\newpage
\begin{center}
{\bf Figure captions}
\end{center}

Figure 1: Drift velocity versus field for different impurity
concentration (a) in a narrow band superlattice ($\Delta=20$meV)
and (b) in a broad band superlattice ($\Delta=150$meV). The unit
of impurity concentration $c=1/(464\rm \AA)$. Inset of (a) shows a
clear view of the velocity-field curves for the high impurity
densities $4c$ and $8c$ because the drift velocity is very small.
The impurity scattering strength $V_0=0.1$eV, and the impurity
size $a=20\rm \AA$.

Figure 2: The calculated DOS for different impurity concentrations
for the broad bandwidth superlattice ($\Delta=150$mV).
$c=1/(464\rm\AA)$ and the electric field $F$ is in units of
$10^7$V/m. The zero-field DOS is shown for comparison (dashed
line).

Figure 3: Different effects of impurity strength and concentration
on electron-phonon resonance peaks in a (a) narrow bandwidth and
(b) broad bandwidth superlattice. The parameters used are
$c=1/(464\rm \AA)$, $V_0=0.02$eV, and $a=100$\AA.

Figure 4: velocity-field curves for different impurity sizes with
a fixed $V_0a$ ($=2\rm eV\cdot$\AA) in a (a) narrow bandwidth and
(b) broad bandwidth superlattice. The impurity concentration
$c$=1/(464\AA).

Figure 5: The drift velocity-field relations for different
impurity sizes while $V_0a$ keeps constant ($=0.12\rm eV\cdot$
\AA) in (a) narrow bandwidth and (b) broad bandwidth
superlattices. The impurity density $c=1/(19\rm \AA)$.

Figure 6: (a) calculated DOS for different $ac$, for
$V_0\sqrt{ac}=\rm const.(1.7meV)$, (b) Bryksin-type DOS
($\hbar\sqrt{U}=1.7\rm meV$); and Wacker-type DOS, with a constant
self-energy $\Gamma/2=\hbar\sqrt{U}=1.7\rm meV$.

Figure 7: Reproduction of Bryksin's results for two bandwidths
($\Delta=\rm 20meV,$ $\rm 36meV$). Bryksin's parameter
$\hbar\sqrt{U}=1.7 \rm meV$. Our parameters are $V_0=0.425$meV and
$ac=16$.

\newpage

\newpage

\newpage

\newpage

\newpage

\newpage

\newpage


\begin{thebibliography}{99}
\bibitem{fourteenband} The 14-band methodology is an extension
of the more commonly employed 8-band one to include the
higher-lying antibonding p states perturbatively. Input parameters
for this empirical method are energy levels, matrix elements, and
effective masses of the bulk constituents, and valence band
offsets between layers. The employed parameters are mainly
extracted from Ref.~[\onlinecite{made82}] and are fully reported
in Ref.~[\onlinecite{flatt96}].
\end{thebibliography}
\end{document}